\documentclass[structabstract,printer]{aa} 
\usepackage{graphicx}
\usepackage{natbib}
\usepackage{amsmath}
\usepackage{txfonts}
\usepackage{color}

\newcommand{\be}{\begin{eqnarray}}
\newcommand{\ee}{\end{eqnarray}}

\newcommand {\nbodypp}{\textsc{\mbox{nbody6\raise.4ex\hbox{\tiny++}}}}

\newcommand {\Msun} {\mbox{M$_{\odot}$}}

\begin{document}

\title{Early evolution of the birth cluster of the solar system}
\author{Susanne Pfalzner\inst{1,}\inst{2}}
\institute{
\inst{1}Max-Planck-Institut f\"ur Radioastronomie, Auf dem H\"ugel 69, 53121 Bonn, Germany\\
\inst{2}Kavli Institute of Theoretical Physics, Kohn Hall, University of California, Santa Barbara CA 93106-4030, USA\\ 
\email{spfalzner@mpifr-bonn.mpg.de}}
\date{ }

\abstract
   { The solar system was most likely born in a star cluster containing at least 1000 stars. It is highly probable that this cluster environment influenced various properties of the solar system like its chemical composition, size and the orbital parameters of some of its constituting bodies. }
   {In the Milky Way, clusters with more than 2000 stars only form in two types - starburst clusters and leaky clusters - each following a unique temporal development in the mass-radius plane. The aim is here to determine the encounter probability in the range relevant to solar system formation for starburst or leaky cluster environments as a function of cluster age. }
   {N-body methods are used to investigate the cluster dynamics and the effect of gravitational interactions between cluster members on young solar-type stars surrounded by discs. }
   {Using the now available knowledge of the cluster density at a given cluster age it is demonstrated that in starburst clusters the central densities over the first 5Myr are so high (initially $>$ 10$^5$ \Msun pc$^{-3}$) that hardly any discs with solar system building potential would survive
 this phase. This makes a starburst clusters an unlikely environment for the formation of our solar system.  Instead it is highly probable that the solar system formed  in a leaky cluster (often classified as OB association). 
It is demonstrated that an encounter determining the characteristic properties existing in our solar systems most likely happened very early on ($<$ 2Myr) in its formation history and that after 5Myr the likelihood of a solar-type star  experiencing such an encounter in a leaky cluster is negligible even if it was still part of the bound 
remnant. This explains why  the solar system could develop and maintain its high circularity later in its development.}
 {}

\maketitle

\section{Introduction}
The average stellar density in the solar neighbourhood is 0.112 stars pc$^{-3}$, meaning that the Sun resides in relative isolation. This might not have been the case at the time of its formation.
Most stars form within groups or clusters of stars (Lada \& Lada 2003, Porras et al. 2003, Evans et al. 2009). The majority of these clusters and associations dissolve within some tens of Myr (Lada 2006, Portegies-Zwart et al. 2010). Even the long-lived open clusters do not exceed a lifetime of some hundreds of Myr. At the current epoch the Solar System is about 4.6 Gyr old. This means if the Sun formed in a cluster environment, the solar birth cluster would have dissipated a long time ago (Portegies-Zwart 2009).  

Several characteristics of the today's solar system give clues to the early formation process.  These are

\begin{itemize}
\item the outer boundary of the planetary system at 30 AU and inner boundary of the Kuiper belt at $\sim$ 45 AU.
\item the much larger eccentricities and inclination angles of the Kuiper Belt objects, as a group, than the planetary orbits
\item the presence of precursors to the short-lived radioactive isotopes  $^{60}$Fe and  $^{26}$Al in meteorites 
\end{itemize}

These properties strongly indicate that the birthplace of the sun is likely to have been a cluster environment \citep{gaidos:95,fernandez:97,adams:10}. Perhaps the strongest constraint comes from the presence of the short-lived radioactive isotopes $^{60}$Fe and  $^{26}$Al inferred from their precursor products in meteorites \citep{wasserburg:06,thrane:06,wadhwa,duprat:08,dauphas,gritschneider:11}. These isotopes have half-lives of  2.62 Myr and 0.717Myr, respectively. Since they decay so quickly, their presence in meteorites means that they must have been  incorporated into the early Solar System very soon after they were produced. 

A supernova is the most likely source for these short-lived radioactive isotopes, in particular the isotope $^{60}$Fe is extremely difficult to produce otherwise than by stellar nucleosynthesis 
(Looney et al. 2006, Williams \& Gaidos 2007, Gounelle et al. 2009). A range of supernovae progenitor masses $M_{sp}$ are possible, and no mass scale produces perfect abundances. However, these studies suggest that stars with $M_{sp} \approx$ 25\Msun\ provide the best fit to the ensemble of short-lived radioactive nuclei (Adams 2010, and references therein). 
Stars of such high mass usually form only in relatively massive clusters. External enrichment of short-lived radio-isotopes suggests a cluster with at least $N >$ 4000 in order to have a reasonable chance of producing a star with $M_{sp}$=25\Msun  (Lee et al. 2008).

The relatively high concentration of these short-lived nuclei means not only that they should have been formed just after the formation of the Sun but also fairly close to it. If not, the 
cross-section of the solar protoplanetary disc would have been too small to account for the amount of material. Several authors have estimated the maximum distance of the solar system when this supernova exploded, obtaining values between 0.2 and 2 pc (for a discussion, see Adams 2010).

The outer boundary of our planetary system (30AU) and inner boundary of the Kuiper belt at 45 AU are likely significant for the early history of the solar system, too. This outer edge is relatively sharp - as best illustrated by the low mass contained in the total of the large number of Kuiper belt objects, which is only $\sim$ 0.01 - 0.1 times the mass of the Earth (Bernstein et al. 2004). The protoplanetary disc from which the solar system eventually formed was most likely  considerably larger at the time of planet formation. Possible processes that could have led to the truncation of the disc include: gravitational interactions with other cluster members \citep{ida:00,kobayashi:01,pfalzner:05,kenyon:04,spurzem:09}, photo-evaporation by nearby massive stars \citep{adams:06,owen:10,mitchell:11} or the supernova explosion itself (Chevalier 2000). Each of these scenarios require a cluster environment of relatively high stellar density.
 
An alternative explanation for the sharp edge in the mass distribution in the solar system would be the Nice model (Gomes et al. 2005), which suggests that the migration of the giant planets caused a planetesimal clearing event leading to a late heavy bombardment (LHB) at 880 Myr. However, Booth et al. (2009) find that collisional processes are important in the Solar system before the LHB and that parameters for weak Kuiper belt objects are inconsistent with the Nice model interpretation of the LHB. 

The much larger eccentricities and inclination angles of the Kuiper Belt objects as a group compared to the planetary orbits, also point to an encounter being the most likely cause.  The most prominent example is Sedna with a periastron of 76AU, an aperastron of 937AU and an eccentricity of 0.8527. It is well beyond the reach of the gas giants and could not be scattered into this highly eccentric orbit from interactions with Neptune alone \citep{gomez:05}. The most straightforward explanation for these high eccentricities would be some type of dynamical interactions (Morbidelli \& Levison 2004) - either a close single encounter or a wide binary solar companion (Matese et al. 2005).

\begin{table}
\caption{Constraints on the properties of the solar birth cluster, where $N$ is the number of cluster members and $\rho_c$ is the central cluster density.}
\begin{center}
\begin{tabular}{lll}
variable &  value & limiting factor  
\\[0.5ex]
\hline
\\[-2ex]
$N$              & $>$ 4000       &   chemical composition  $^1$                \\[0.5ex]
$N$              & $<$ several 10$^4$   &   radiation field  $^2$                               \\[0.5ex]
$\rho_{c}$    &  $>10^3$ \Msun pc$^{-3}$   &   Sedna orbit  $^3$   \\[0.5ex]
$\rho_{c}$    &  $<10^5$ \Msun pc$^{-3}$   &   Sedna orbit  $^4$  \\[0.5ex]
\end{tabular}
\label{table:models}
\end{center}
Note:$^1$Lee et al. (2008),  $^2$Adams (2010),   $^3$Brasser et al. (2006),  $^4$Schwamb et al. (2010).
\end{table}

If an encounter is responsible for the disc cut-off and the Sedna orbit, then it must have taken place
when the stellar density was still relatively high. It is generally assumed that clusters expand significantly after gas expulsion which is accompanied by a dramatic drop in stellar densities. The time scale of this cluster expansion was until recently not well known. The encounter must have occured early on in the history of the solar system \citep{malhotra:08} as otherwise the dynamical interactions due to passing stars would have not only lead to the truncation of the disc and the high eccentrcities in the Kuiper Belt objects, but would also have perturbed the  orbits of the planets, stripped the comets away and led to portions of the Inner Oort Cloud becoming unstable \citep{gaidos:95,adams:01b}. The fact that all of the planetary orbits are nearly in the same plane (inclination angles $<$ 3.5$^\circ$), their low orbital eccentricities($<$0.2) and the presence of the Inner Oort Cloud all point to a lack of severe disruption after the solar system was fully formed. This provided an upper limit on the stellar density in the solar birth cluster for ages $>$ 30Myr.

Recently a number of authors tried to translate constraints derived from the meteorite composition, the disc cut-off and the Sedna orbit into properties of the solar birth environment.
These are basically limitations on the number of stars $N$ in the solar birth cluster and its central density $\rho_c$. For a summary of their results see Table 1.  These conditions do not necessarily need to be fulfilled over a long period of time. So the condition on $N$ based on the chemical composition needs only to be fullfilled before the supernova explodes, whereas the upper limit based on the radiation field needs to be fulfilled after gas expulsion from the cluster - before gas expulsion photo evaporation is much less efficient - until about 5 Myr, when discs are anyway largely dispersed (Hernandez et al. 2007). The conditions on the cluster density must only be fulfilled at the instant of the encounter.

Most investigations conclude that a membership of $N \approx$  10$^3$ to several 10$^4$ is the most  likely scenario for the solar birth cluster. The lower limit of this estimate is based on the demand that the solar birth cluster should have contained enough stars in order to that there is a reasonable likelihood
that it hosted the required supernova progenitor (for details see Adams 2010). Chemical considerations also suggest that the Sun formed in the presence of strong FUV radiation fields, where rough estimates indicate a birth cluster with $N >$ 4000 (Lee et al. 2008).
Really massive clusters with $N>$10$^5$ have been considered as birth place of the solar system (Williams \& Gaidos 2007, Hester et al. 2004) too. However, the radiation fields provided by such massive young clusters are potentially disruptive. Nevertheless, the early solar nebula could survive in clusters with $N \sim$ several 10$^4$ provided it spends enough time in the outer cluster regions(Scally \& Clarke 2001, Mann \& Williams 2009).  

The cut-off radius of the disc is also used to constrain the membership of the solar birth cluster.
The basic method here is to determine the required periastron distance $r_{peri}$ of the Sun with respect to the star encountered to obtain a cut-off in the disc. In a second step this is related to a certain cluster membership.
The lower limit of $r_{peri}$  is derived from the results of \cite{hall:96} and \cite{kobayashi:01}, which state that parabolic, prograde encounters with solar-type stars lead to a reduction of the disc size to approximately 1/2 - 1/3 of the periastron distance. Assuming a typical disc size of 100AU, the cut-off radius translates into a periastron distance of $r_{peri} \approx$ 100AU to a grazing encounter.   Here a number of simplifying assumptions (parabolic,solar-type star) are made which might require reconsidering \citep{olczak:08,olczak:10}.  Other authors \citep{morbidelli:04,looney:06} come to somewhat different periastron values. So to cover the entire parameter range, we consider periastra from 100 AU to 1000AU to be potentially solar-system forming encounter events.  This translates into a solar birth cluster in the approximate range $N$ = 10$^3$- 10$^4$ (Adams 2010). 

To recap, the abundances of short-lived radioactive isotopes and the cut-off at 30AU both strongly hint at a solar birth cluster size in the range of 4000 $<N<$ 10$^5$. Assuming that the stellar masses in the cluster follow the initial mass function (IMF) with an average stellar mass of $m_{av}=0.5 \Msun$ this is equivalent to a total cluster mass $m_{cl}$ of 2000 \Msun $<m_{cl} < $ 0.5 $\times$ 10$^5$\Msun. 

What about the stellar densities in these clusters?  
The density of the solar system birth cluster is so far mainly constrained by considering what kind of encounters are necessary to result in a Sedna-like orbit. It is a point of debate whether there is an encounter scenario that can account for both features - cut-off and the Sedna orbit - resulting from a single encounter, or whether two separate encounters are necessary.

Performing investigations of the origin of Sedna's orbit,  Brasser et al. (2006) required {\em central} cluster densities higher than 10$^3$ \Msun pc$^{-3}$ whereas \citet{schwamb:10}  ruled out central densities $\geq$10$^5$ \Msun pc$^{-3}$. Both worked with somewhat different periastri than Adams(2010), which can be summarized in that an encounter between 100 - 1000 AU was required to result in the high eccentricities of the Kuiper belt objects. Brasser et al. (2006) and Schwamb et al. (2010) give limits to the {\em central}, and not the {\em mean} density of the cluster, because due to mass segregation the most massive stars of a cluster are usually located close to the cluster center. Since the solar system must have formed close to a supernova which had a massive star as progenitor,  the Sun most likely resided close to the cluster center. 

In summary, it can be said that the presence of short-lived isotopes, the fall-off of the mass distribution of the solar system and orbit of Sedna all combine to constrain the solar birth cluster to a membership of a few thousands to ten-thousands of stars with a central density of 10$^3$ $\leq \rho_c \leq$ 10$^5$ \Msun pc$^{-3}$.

Recently it was found that massive clusters evolve in specific ways after gas expulsion (Pfalzner 2009). In this paper we investigate how this knowledge can be applied to deduce the birth environment of the Solar system. Section 2 recaps the temporal development of massive clusters.
In Section 3 we show by means of simple approximations that the Solar system is unlikely to have developed in a star burst cluster. Section 4 shows the results of detailed modelling of the temporal
developement of solar type star-disc systems in leaky clusters. The limitations and consequences
of these results are discussed in Section 5.

\section{Cluster environments}

The requirement of the mass of the solar system birth cluster to be in the range 2000 \Msun $<m_{cl} < $ 5 $\times$ 10$^4$\Msun\  puts it in the class of massive clusters.
In the Milky Way massive young star clusters (t$_c<$ 20 Myr, $M_c >$ 10$^3$ \Msun) are observed to have {\em mean} densities ranging from less than 0.01 to several 10$^5$ \Msun pc$^{-3}$ and radii from 0.1pc to several tens of pc. This means clusters with properties, in terms of mass and density, demanded for the solar birth cluster exist also at the present time in the Milky Way. 

\begin{figure}
\begin{center}
\includegraphics*[width=0.5\textwidth]{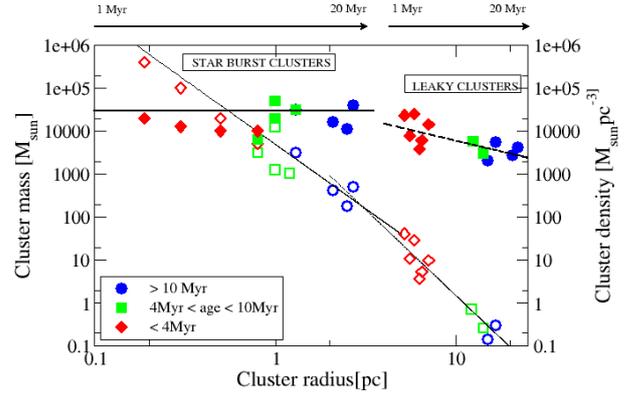}
\caption{Observed cluster mass (filled symbols) and derived density (open symbols) as a function of cluster size for clusters more massive than 10$^3$ \Msun as published by Pfalzner (2011). The values were taken from Figer (2008), Wolff (2007) and Borissova et al. (2008) and references therein.}
\label{fig:cluster_rad}
\end{center}
\end{figure}

Recently it was demonstrated that such massive clusters evolve in time only along one of two well-defined density-radius tracks - strongly suggesting a bi-modal cluster formation (see Fig.1) (Pfalzner 2009). 
The two cluster groups differ considerably in their properties:
One group, termed {\it starburst clusters}, comprises young compact clusters which develop from a radius $R_c$ of $\sim$ 0.1pc at 1 Myr to about 1-3pc at 20 Myr. Simultaneously the density drops from the initially very high mean stellar density of 
$\geq$100$^5$ \Msun pc$^{-3}$  to 
10 \Msun pc$^{-3}$ during that timespan. All these starburst clusters are born with the same mean cluster mass, radius and density and then simply diffuse as $R_c^{-3}$ while more or less retaining their mass. The expansion basically proceeds linearly in time, i.e. as R$_c$ $\propto$ t$_c$ with an expansion velocity of  0.1-0.2 pc/Myr.

The other group, termed {\it leaky clusters},  contains many clusters that are usually termed OB associations. This group consists of clusters covering the same age span, but having on average larger radii and lower mean cluster densities ($<$1 - 10$^3$  \Msun  pc$^{-3}$).  Starting with cluster radii of $\sim$2pc at 1 Myr, they also expand in a time sequence, however, here the cluster 
density decreases as R$_c^{-4}$ rather than R$_c^{-3}$.  In these clusters in addition to diffusion, the clusters lose mass. The cluster size does not increase linearly with time, but approximately as t$_c^{0.6-0.7}$ and reaches about 25pc at 20 Myr. 

This restriction of the development of clusters more massive than 10$^3$\Msun\
along only two possible developmental tracks has profound consequences on the formation history of our solar sytem: The solar system  inevitably developed either in a starburst or a leaky cluster. The now available knowledge of how the average cluster density $\rho_c$ changes with cluster age $t_c$ in these two environments,  reduces the scenarios of the solar birth cluster to {\em only two possible developmental paths}. It now remains to determine which of the two environments - starburst cluster or  leaky cluster -  is the more likely origin of the solar system.

\section{Solar system in a starburst cluster?}

Starburst clusters retain more or less their complete mass over the first 20 Myr. They are likely to remain a bound entity until the galactic tidal field disrupts them. This probably only happens for ages $t_{sb}>$100Myr. A statistical argument against starburst clusters as solar birth hosts would be that probably only $\sim$10\% of the stellar population is born within systems that remain gravitationally bound over timescales longer than 100 Myr (e.g., Battinelli \& Capuzzo-Dolcetta 1991, Adams \& Myers 2001). Most stars are born within clusters that dissolve quickly, after only a few tens of Myr
(Lada \& Lada 2003, Porras et al. 2003).

\begin{figure}
\includegraphics[width=0.5\textwidth]{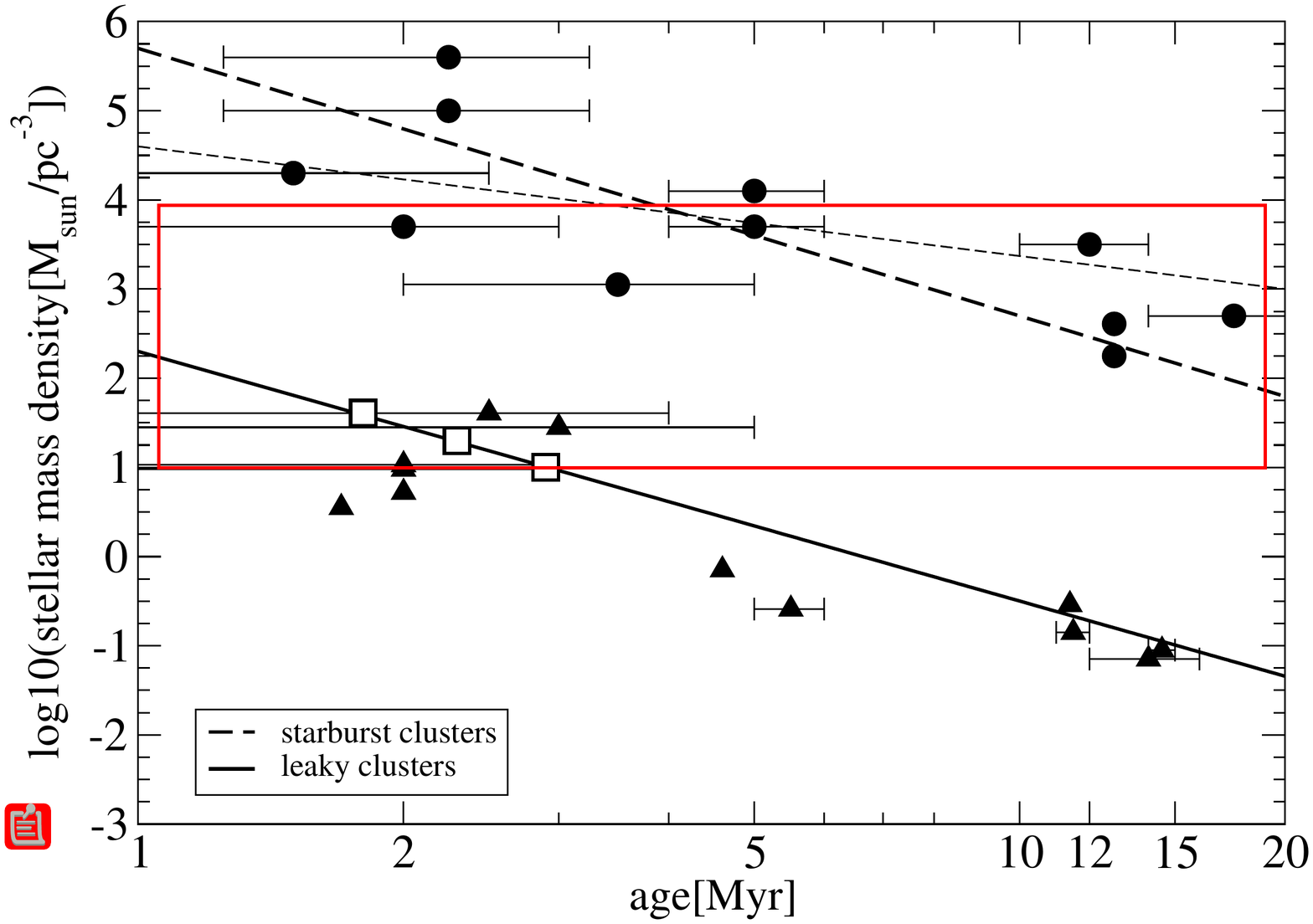}
\caption{Mean cluster density as function of the cluster age for starburst (circles - observed values and dashed line - interpolation as in Pfalzner 2009) and leaky clusters(triangles and solid line, observations and interpolation, respectively).  The red box shows the density requirement for solar-system forming encounters to be likely. The  open squares show the parameters of the clusters modelled in chapter 4.  The thin dashed line shows a fit through the data case where the Arches cluster is not considered in the starburst cluster sequence. Here only the errors in cluster age are given, the errors in cluster densities can be found in Fig. 2 of Pfalzner 2009.}
\label{fig:dens_age}
\end{figure}

However, the strongest argument against starburst clusters is their very high initial density of $\rho_{sb}^i>$10$^5$ \Msun pc$^{-3}$, combined with the high number of massive stars.
From the density-size and size-age relations derived in Pfalzner (2009), it follows that the development of the mean stellar density $\rho_{sb}$ in a starburst cluster environment can be approximated by (see also dashed line in Fig. 2)
\be \rho_{sb} \approx 5\times 10^5 t_c^{-3 \pm 0.3} [\Msun \mbox{pc}^{-3}].\ee 
where and in the following $t_c$ is in units of Myr.
Before comparing this density development in starburst clusters with the requirements derived by Brasser et al. (2006) and Schwamb et al. (2010) for the solar birth cluster, one has to correct for the fact that the values by Brasser et al. and Schwamb et al.  are {\em central} densities whereas Eq. 1 describes the mean density. The reason why they consider central cluster densities is that many star clusters exhibit mass segregation at early ages ($<$ 1Myr) with the most massive stars preferrentially being located close to the cluster center. From that it follows that most likely 
the Sun was located near the cluster center as otherwise it could not have been in close vicinity to the exploding supernova.  

 The mass density values quoted in Pfalzner (2009) are obtained by dividing the cluster mass by its volume - implying an average density. The stellar density in a cluster is in fact a strong function of the radial distance to the cluster center. Consequently the central density of a cluster is usually much higher than its mean density. For model clusters of the Orion Nebula Cluster (details of the simulation method can be found in Olczak et al. 2010) we find - depending on the definition of `central' - the central cluster density ranges from a factor 5-100 times higher than the average stellar density obtained at the half-mass radius.
%
%
%
%
%
Observations confirm this fact at least for low-mass clusters: for example, Guthermuth et al.(2003) found that in all of the three embedded clusters that they observed, the mean volume densities were of the order 10$^2$-10$^3$ \Msun pc$^{-3}$, whereas their peak volume densities ranged from  10$^4$-10$^5$ \Msun pc$^{-3}$. Whether this holds as well for high-mass clusters needs further
investigation.

So the requirement by Brasser et al. (2006) and Schwamb et al. (2010) that the central density in the solar system birth cluster should lie in the range 10$^3$\Msun pc$^{-3} < \rho_c^{sb} < $10$^5$\Msun pc$^{-3}$ translates into a mean density of 
10\Msun pc$^{-3} < \rho_m^{sb} < $ 10$^4$\Msun pc$^{-3}$. In Fig. 2 this area is depicted by a red box. It can be seen that the density requirement for the solar birth cluster overlaps with the development of starburst clusters at ages $t_c >$ 4 Myr. 

However, at younger ages starburst clusters have an up to 10$^2$ times
higher density. At such high densities many of the stars would be completely stripped of their discs due to gravitational interactions \citep{olczak:10}.

\begin{figure}
\includegraphics[width=0.5\textwidth]{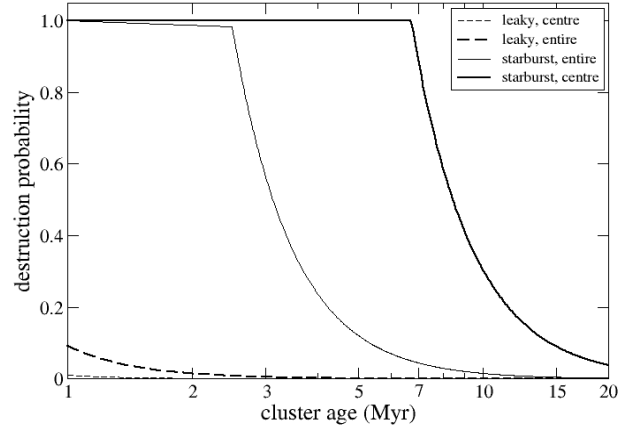}
\caption{Probability for encounters closer than 100AU taking place as function of cluster age. The solid and the dashed lines show the estimates for starburst and leaky clusters, respectively. The thick and thin lines distinguish between the entire cluster and the central area only.}
\label{fig:dens_age}
\end{figure}

The average time $\tau_{enc}$ until a star undergoes an encounter with a certain periastron distance $r_{peri}$ can be approximated by \citep{davies:11}

\be \tau_{enc} \simeq \mbox{33}\mbox{Myr}\left( \frac{\mbox{100pc}^{-3}}{n}\right) 
                                                         \left( \frac{v_\infty}{\mbox{1km/s}}\right) 
                                                         \left( \frac{\mbox{10}^3\mbox{AU}}{r_{peri}}\right) 
                                                         \left( \frac{\Msun}{m}\right), \ee

where $n$ is the stellar number density in pc$^{-3}$, $v_\infty$ the mean relative velocity at infinity of the cluster stars and $m$ the mass of the considered star in \Msun.  With the cluster mass $M_c \approx$ 0.5$ N_c$ \Msun, where $N_c$ is the number of cluster members, it follows that $\rho [\Msun pc^{-3}] \approx$ 0.5$ n$. Therefore
the encounter probability in a star burst cluster can be approximated by

\be \tau^{sb}_{enc} \simeq \mbox{33}\mbox{Myr}\left( \frac{\mbox{200 \Msun}}{pc^{-3}}{\rho_{sb}}\right) 
                                                         \left( \frac{v_\infty}{\mbox{1km/s}}\right) 
                                                         \left( \frac{\mbox{10}^3\mbox{AU}}{r_{peri}}\right) 
                                                         \left( \frac{\Msun}{m}\right), \ee  

In a simplified picture an encounter closer than 100AU would lead to a truncation of the disc to $r_{disc} <$30AU (Adams 2010). This deprives the disc of the material necessary
 to form the outer planets, so the resulting planetary system  would no longer resemble our solar system.  Equation 3 can be used to estimate the likelihood of such a destructive encounter to occur in a starburst environment. Fig. 3 (solid lines) shows the likelihood of a destructive encounter ($r_{peri} < $100AU)
in a starburst cluster as a function of cluster age using the density development given by Eq. 1. The thin line shows the probability for such an encounter to happen in the entire cluster and the thick line the result for the cluster center, assuming the density there to be on average about 20 times denser then the mean cluster density. Here disc destruction is assumed if each star is likely to experiences at least one encounter closer than 100 AU during a 1 Myr  time interval. Or in other words, the destruction probability is 1, if $\tau^{sb}_{enc} < $ 1Myr and 1/$\tau^{sb}_{enc}[Myr]$ for $\tau^{sb}_{enc} > $ 1Myr. This definition does not include repeated encounters. So in reality disc destruction would be even higher than indicated in fig. 3.

 In both cases the cluster environment is extremely collisional and encounters closer than 100AU are very common. For the outskirts of the cluster this changes after 3 Myr but for the central area continues up to 7-8Myr. Disc destruction happens on fairly short timescales.
For example, for a solar-type star in a cluster with $\rho_c>$10$^5 \Msun$ pc$^{-3}$, which is typical for starburst cluster environments younger than $<$ 2Myr, an encounter closer than 100AU can be expected already after $\tau_{d}^{sb} \ll $0.3Myr. 
The disc destruction happens on such short timescales ($<$1Myr) that it would be difficult to form a planetary system before the disc is destroyed. Though one can not completely exclude the possibility of solar system having developed in a starburst cluster environment, it seems rather unlikely. Moreover, photo-evaporation\citep{adams:10}, which has not be included in this estimate, would lead to additional disc destruction.

Recent observations by Stolte et al. (2010) suggest that there exist discs around some of the stars in the starburst cluster Arches, however, the vast majority is devoid of inner discs. They only find discs around B-type stars with a frequency of 6\% $\pm$ 2\%. For solar-type stars the likelihood of retaining a disc would be even lower (Pfalzner et al. 2006). Assuming for a moment that the solar system happened to develop in one of such rare cases of its disc surviving the first Myr, the chances of  discs surviving even longer would actually increase, because starburst clusters become much less dense as they age. Nonetheless, at an age of 20 Myrs the average stellar density would still be several 100 \Msun pc$^{-3}$. This is such a high density that, even if one assumes that some discs in a starburst environment survive long enough to form a planetary system, it would inevitably lead to a strong perturbation when the planetary system is close to being completed. Interactions with other cluster members would inevitably lead to disturbances destroying the near circularity and co-planarity of the solar systems. 

The only way that our solar system could have formed as part of a starburst cluster would be if it had formed at the outer edge of this cluster, always remained there and never transversed the cluster center. A rough estimate using Eq. 2 would result in the requirement that the solar system would then have to have been located at 4-5 times the half-mass radius at the time of cluster formation. 

It has been suggested that Arches might be exceptional for Milky Way clusters as it is thought to evolve quite differently from the clusters in the disk because of the unusually strong tidal fields (only $\sim$30 pc away from the Galactic center). If we exclude this cluster from our investigation, this means that the size of the cluster at the time of gas expulsion could be larger than anticipated by including it. The thin dashed line in Fig. 2 shows the density developemnt excluding Arches. The cluster expansion starts at lower densities on the developmental track in Fig. 1. However, even in this case the cluster would go through a phase of 3-4 Myr where basically every star in the cluster center undergoes at least one encounter closer than 100 AU encounter in every 1 Myr timespan. So it still holds that nearly all discs are likely destroyed in the cluster center. However, there is an increased change that discs could survive in the cluster outskirts. The requirement reduces to the condition that the solar system would then have to have been located at 2-4 times the half-mass radius at the time of cluster formation.

However, at the same time a massive star would have to have been at these outskirts of the cluster to account for the observed chemical composition. Although this scenario cannot be excluded, it seems fairly unlikely.  In the following we will see that it is much more likely that the solar system developed in leaky cluster environment.

\section{Solar system in a leaky cluster}

For the leaky clusters the age-dependence of the average cluster density $\rho_l$  can be approximated by \citep{pfalzner:09}
\be \rho_{l} \> = 150 t_c^{-2.6 \pm 0.2} [\Msun \mbox{pc}^{-3}]  \hspace{1cm} \mbox{for } t{\ge} \mbox{ 1 Myr}\ee 
where $t_c$ is in units of Myr. This is illustrated by the solid line in Fig. 2.
Early on in the cluster development the mean cluster density will be approximately a few times 10 \Msun/pc$^3$ to 100\Msun/pc$^3$ and will rapidly decrease over the following 20 Myr.  Looking again at Fig. 3 (dashed lines) one sees that disc destruction approximated from 
\be \tau^{l}_{enc} \simeq \mbox{33}\mbox{Myr}\left( \frac{\mbox{200pc}^{-3}}{\rho_{l}}\right) 
                                                         \left( \frac{v_\infty}{\mbox{1km/s}}\right) 
                                                         \left( \frac{\mbox{10}^3\mbox{AU}}{r_{peri}}\right) 
                                                         \left( \frac{\Msun}{m}\right), \ee
is not an issue in the leaky cluster environment. Even early on in the cluster development, less than 10\% of the stars in the cluster center are likely to undergo encounters closer than 100AU and in the entire cluster it $<$ 1\% of solar-type stars that are affected.

Fig. 2 shows that the density requirement for the solar birth cluster overlaps with the development of leaky clusters for ages $t_c <$ 4 Myr. At later times the cluster density is considerably lower making a solar system forming encounter unlikely. 
This is the central result of this study: with a very high probability,  {\em the solar system developed in a leaky cluster.}

This development of the stellar density in such leaky clusters also neatly solves the seemingly contradictory situation that a dense birth cluster is required to provide for the nuclear enrichment and explain Sedna's orbit, but at the same time, a less interactive environment is needed to avoid disruptive dynamical interactions with the formed solar system.

In the following we want to go beyond the simple estimates of Eqs. 2-4 and determine what the development of the solar system in a leaky cluster means in terms of its history by performing numerical simulations of its encounter dynamics.

\subsection{Numerical method and  initial conditions}

The clusters identified in Fig.1 as part of the leaky cluster sequence are all exposed clusters, where the gas has been expelled from the cluster.  Although gas expulsion seems to be a vital ingredient for the cluster expanding in the observed way, there are currently many unknowns 
setting the course of the following cluster dynamics and mass loss in the leaky cluster sequence. These uncertainites concern the
time scales of the gas expulsion, the dynamical state - equilibrium, sub -or supervirial - before gas expulsion and to what degree early substructuring plays a role.

Therefore instead of trying to include these unknown processes here, existing simulations (Olczak, Pfalzner \& Eckart 2010) of clusters in virial equilibrium at different densities are used to obtain a first estimate of the dynamics of the early solar environment  as a function of cluster age.    The densities of the modelled clusters are indicated in Fig. 2 as open squares. The densities of these model clusters decrease very slightly over the considered timespan of 1 Myr. 
This means the observed cluster density development is replaced by snapshots of clusters of fairly constant density. The actual numerical method and the model set up is described in detail in Olczak, Pfalzner \& Eckart (2010).
  
The main aspects of the model can be summarized in the following way: 
All simulations were performed using the direct many-body code Nbody6++  (Aarseth 2003; Spurzem 1999).  For simplicity it was assumed that all stars are initially single and no primordial binaries are considered. This seems to be justified, as  Adams et al. (2006) and Brasser et al. (2006) found in their investigation of potential solar system environments that binary scattering events do not influence the results significantly. The effect of stellar evolution has not been included. This is justified by us modelling only the first 5 Myr of the cluster development in detail, where stellar evolution plays a minor role.

The stellar masses in the cluster were  sampled from the initial mass function given by Kroupa(2001). The cluster members cover the mass range from the lower mass end of 0.08\Msun\ to the currently accepted upper mass limit of 150 \Msun (Koen 2006; Oey \& Clarke 2005; Ma{\'{\i}}z Apellaniz et al. 2007; Zinnecker \& Yorke 2007; Weidner \& Kroupa 2006; Figer 2005). The velocity distribution is Maxwellian and the cluster is assumed to be in virial equilibrium.
In reality leaky clusters at that phase of their development are actually supervirial as the cluster still adjusts after the gas explusion process. Bastian \& Goodwin (2006) find for somewhat larger clusters  than we study here and with star formation efficiencies of $\approx$ 30\%, that it takes several 10$^7$ years until the clusters reach their new equilibrium state.  However, not really knowing the gas expulsion dynamics and star formation efficencies in leaky clusters in detail, we resort to
the easiest assumption namely treating the virial equilibrium case. We will discuss later in how far this might influence our results.

The cluster density profile was chosen in such a way that it resembles the observed ONC profile \citep{Mc02} after 1Myr of simulation. We chose an ONC-like density profile because it is the best known massive cluster just before gas expulsion. As we model the early stages of exposed clusters we assume that the shape of the density profile does not alter significantly (at least in the central areas) over the timespan of the first $\approx$ 5 Myr. From much older open and globular clusters it is known that they have less centrally condensed Plummer-type profiles, but this development happens on timesclaes of $>$100 Myr to Gyr.

 Due to the adopted stellar number density distribution, which is roughly represented by $\rho = \rho_0 r^{-2}$ , the density of the models scales as the stellar number, $ N = \int_0^R \rho(r) r^2 dr d\Omega \propto \rho_0 R.$. Different initial stellar densities are obtained by varying the stellar numbers to 1000, 2000, 4000 (ONC), 8000, 16 000, and 32 000 (see Table 2). These density-scaled cluster models have been simulated with the same initial size of R=2.5 pc corresponding to the size of the Orion Nebula Cluster (ONC).  The central density in the different models covers the range of   $\approx$1-42$\times$10$^4$pc$^{-3}$. 

For each cluster model a set of simulations has been performed with varying random configurations of positions, velocities, and masses, according to the given distributions, to minimise the effect of statistical uncertainties. 
For the clusters with 1000, 2000, 4000, 8000, 16 000, and 32 000 particles, a number of 200, 100, 100, 50, 20, and 20 simulations, respectively, were performed to ensure statistical robustness. Due to the relatively high particle number  in the simulation the encounter statistics of the stars close the center is not influenced by the change in simulation particles (see Olczak et al. 2010). Therefore we can treat the obtained encounter probabilities as direct function of the cluster densities.  

By contrast to previous studies we do not consider the simulation of clusters of different densities as different possible solar system birth environments, but as consecutive stages
in the development of a leaky cluster environment which we can assign to certain cluster ages. In the following this is described in more detail.

\begin{table}
\caption{Cluster models, resulting mean cluster densities and age, where these densities are reached in the leaky cluster development. The last column shows the probability for a solar-type star located within 0.7 pc from the center to ungergo an encounter between 100AU and 1000AU}
\begin{center}
\begin{tabular}{llll}
No. of stars &  mean density & cluster age  & encounter 
\\[0.5ex]
                   &  \Msun / pc$^3$ & Myr        & probability
\\[0.5ex]
\hline
\\[-2ex]
$32 000$              &  4.2 $\times$ 10$^4$    &   -  & -          \\[0.5ex]
$16 000$              &  2.1 $\times$ 10$^4$    &   -  & 63.4\%  \\[0.5ex]
$  8 000$              &  1.0 $\times$ 10$^4$    &   -  & 41.9\%  \\[0.5ex]
$  4 000$              &  5.3 $\times$ 10$^3$    &   1.8 Myr  & 28,4\%  \\[0.5ex]
$  2 000$              &  2.7 $\times$ 10$^3$    &   2.3 Myr  & 21,3\%  \\[0.5ex]
$  1 000$              &  1.3 $\times$ 10$^3$    &   2.8 Myr  & 16,9\%  \\[0.5ex]
\end{tabular}
\label{table:models}
\end{center}
\end{table}

\begin{figure}
\includegraphics[width=0.5\textwidth]{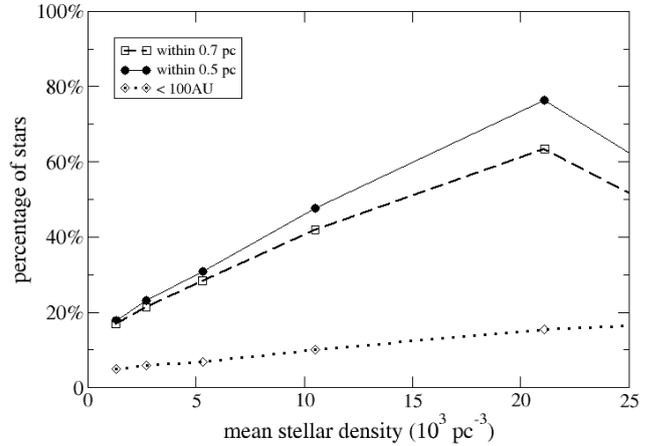}
\caption{Percentage of solar-type stars (0.8\Msun $< M_s <$ 1.2\Msun)  located within 0.7pc (solid line) or 0.5pc (dashed line) from the cluster center at the moment of the encounter which has a periastron in the range 100AU -1000AU during the first Myr of the cluster development as a function of mean cluster density. The mean density is the spatial and temporale mean during this 1 Myr timespan. The dotted line shows the percentage of stars that have  an encounter closer than 100 AU.}
\label{fig:cluster_enc_dens}
\end{figure}

\subsection{Encounter statistics}

While simulating the stellar dynamics of above described clusters, for solar-type stars the parameters (mass ratio, periastron, eccentricity) of every encounter closer 1000 AU is recorded.
Here we define solar-type stars as a stars in the mass-range 0.8-1.2 \Msun\ .

The Sun must have been close to a massive star that exploded as a supernova and enriched the solar system with $^{60}$Fe and  $^{26}$Al. To consider this we again take the ONC as a guide for the typical cluster environment. Here the result of mass segregation is that most massive stars are located within 0.5 pc of the cluster center.  To model the early history of the Solar system we therefore consider only stars within 0.5pc and 0.7pc form the cluster center. The latter denotes the central cluster area of 0.5 pc plus the estimated distance of the Sun to the supernova of 0.2pc.

\begin{figure}
\includegraphics[width=0.5\textwidth]{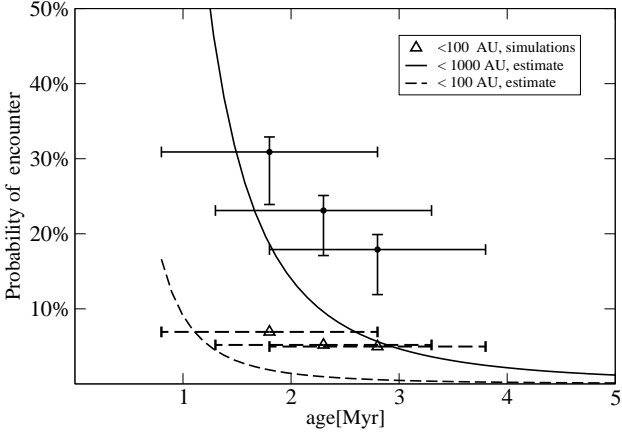}
\caption{The dots show the simulation results of the percentage of solar-type stars (0.8\Msun $< M_s <$ 1.2\Msun) located within 0.5pc from the cluster center which have a single encounter with a periastron in the range 100AU -1000AU as a function of cluster age for leaky cluster environments. The triangles shows the percentage of encounters closer than 100 AU. The solid and the dashed lines show the equivalent estimates according to Eq.2.}
\label{fig:cluster_dens_rho}
\end{figure}

Figure  \ref{fig:cluster_enc_dens} shows the probability of a solar-type star located within  0.5pc (solid line) and 0.7pc (dashed line) from the cluster center having an encounter between 100 AU and 1000AU during 1 Myr of cluster developement as a function of the cluster density.  
Stars that have encounters closer than 100 AU before or after in addition are excluded, because such encounters would truncate the disc to such a degree that they would be too small for developing into a solar-type planetary system. As expected the likelihood of encounters between  100 and 1000AU increases considerably with higher cluster density. However, so does the value of encounters closer than 100AU - the proportion of "destroyed" discs (dotted line in Fig.\ref{fig:cluster_enc_dens} ).  Therefore in very dense clusters ($\rho > $ 2$\times$ 10$^4$) the likelihood of a solar system forming encounter decreases again, because an increasing number of discs have an encounter closer than 100AU that makes it unsuitable for developing into a solar system.

Given the encounter likelihood at different central densities, we can now deduce the
encounter statistcs during the development in a leaky cluster with age using eq. 3.  Table 2
shows how the different densities correspond to ages in the leaky cluster sequence. No densities
above $\sim 5 \times$ 10$^3$ \Msun pc$^{-3}$ are reached in the leaky cluster sequence. However, this does not necessarily mean that the cluster never went through a phase of higher stellar density. Such a situation might have existed in the earlier embedded phase while the cluster was still in its formation process. However, little information about the stellar densities
in forming massive clusters exist and existing knowledge can be interpreted very differently 
(for different views see for example Pfalzner 2011 and Parmentier et al. (subm.)). Therefore we restrain this study to the exposed phase only.

Figure 5 shows the probability for an encounter at a distance of 100 AU to 1000 AU in such an
evolving leaky cluster.   Here table 2 and eq. 4 have been used to equate a given density to a cluster age. The encounter rate is obtained as an average over 0.5 Myr. During that time interval the cluster density decreases faster in the leaky cluster development than in the simulated systems, which are in virial equilibrium. Therefore the encounter rates have to be regarded as upper limits which is reflected by the larger error bars towards lower encounter rates. 

Figure 5 shows that in a leaky cluster environment solar-type stars have  a probability of $\sim$ 30\% of experiencing a potentially solar system forming encounter early on in the development. As the cluster density decreases with cluster age, so does the encounter probability. This plot demonstrates that the highest likelihood for a solar system forming encounter is early on in the cluster development ($<$2Myr). After 5 Myr of cluster development encounters below 1000 AU become extremely rare because the density in the cluster has decreased so much. 

It can be seen that the encounter rate measured in the N-body simulations  (circles) is higher than the one expected from the simple estimate of eq. 5 (solid line). This is because there is gravitational focussing around the massive stars, which themselves are concentrated in the cluster center \citep{pfalzner:06,portegies:11}, which in turn leads to a significantly higher number of encounters. The combination of gravitational focussing and mass segregation is not considered in the estimate of eq. 5. 

These more detailed studies support the idea that disc destruction is less of a problem in leaky clusters. Although the simulation value of disc destruction (triangles) is higher than the estimated one (dashed line), still very few solar-type stars have an encounter that is closer than the threshold of 100 AU.

\begin{figure}
\includegraphics[width=0.5\textwidth]{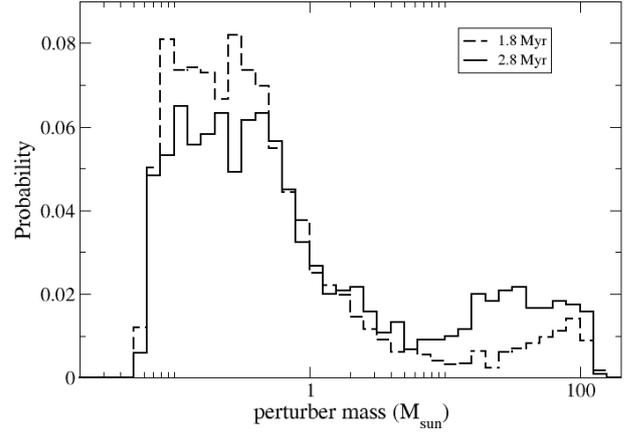}
\caption{Distribution of the perturber mass of the encounters between 100 AU and 1000 AU for solar-type stars in leaky clusters at 1.8 (dashed line) and 2.8 Myr (solid line).}
\label{fig:cluster_dens_rho}
\end{figure}

Figure  6 shows the mass distribution of the encounter partners in a timespan of 1 Myr in these encounters between 100AU and 1000 AU at an cluster age of 1.8 Myr and 2.8 Myr, respectively. It can be seen that there are two maxima in this distribution - one for low-mass and one for high-mass perturbers. Most encounters are with low-mass stars ( $m_2 <$ 0.8\Msun), this is simply due to the fact that they are the most common in the cluster. However, a considerable amount of encounters take as well place with the highest mass stars ($>$ 10 \Msun) of the cluster. As Fig. 6 shows it is actually just as likely for the Sun to have had an encounter between 100 AU and 1000AU with a solar-type star as with a high-mass star despite low numberof the latter. The reason is gravitational focussing. This is most pronounced later in the cluster development when the cluster density is lower. This result shows that the often applied assumption that the perturber was also a solar-type star has to be reconsidered in future work. These new results do not even exclude that the supernova progenitor itself might have been the encounter partner that led to the shape of the solar system.  

\begin{figure}
\includegraphics[width=0.5\textwidth]{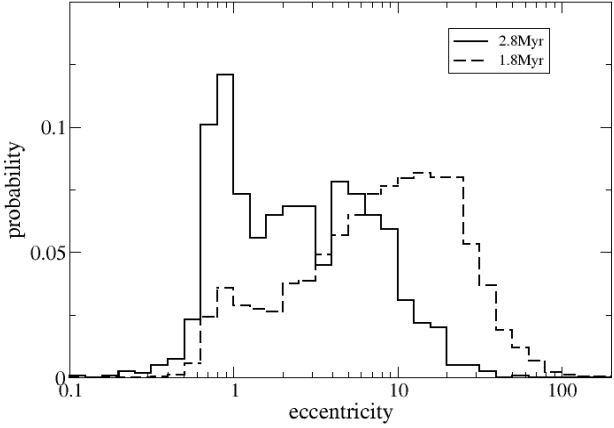}
\caption{Eccentricity distribution of the encounters between 100 AU and 1000 AU for solar-type stars in leaky clusters at 1.8 (dashed line) and 2.8 Myr (solid line).}
\label{fig:cluster_dens_rho}
\end{figure}

There is another point about the solar system forming encounter that might require revision: the eccentricity.
Fig. 7 shows the distribution of the eccentricities of the encounters for two different cluster ages
as the leaky cluster environment develops. In the early cluster development strongly hyperbolic encounters are more likely. Hyperbolic encounters lead to considerably less mass loss in the disc than the equivalent do parabolic ones because the interaction time for disc mass removal is much shorter (see, for example, Pfalzner 2004, Pfalzner et al 2005, Olczak et al. 2010). Only at ages $>$2Myr parabolic encounters start to happen more often. 

The fact that highly hyperbolic encounters dominate in dense cluster environments might increase the chance that the solar system formed in a starburst cluster environment because the disc destruction timescale in these environments might increase. First indications that this might be the case were found by Olczak et al. 2012. However, the problem of the strong radiation field in starburst clusters remains.

In summary, it can be said that the assumption that the perturber that formed the shape of the solar system was a solar-type star on a parabolic orbit is highly questionable (see as well Dukes \& Krumholz 2011). Future investigations will have to take the entire mass spectrum and a much wider eccentricity range into account to narrow down the encounter sceanrio that lead to the formation of the solar system.

\section{Discussion and conclusions}



Previous work puts the cluster population limit of the birth environment of the Sun at 10$^3$ to 10$^5$ members. Recently it was shown that such massive clusters only develop along one of two existing tracks in the density-radius plane - as either starburst clusters or leaky clusters(OB associations). As a consequence the solar system must have developed in one of these two environments. Using the now known temporal density development of starburst clusters, it has been demonstrated here that in starburst clusters the high initial central stellar density ($>$10$^5$ \Msun/pc$^3$)  despite its rapid decline leads almost inevitably to disc truncation over the first 5-10Myr. The resulting discs are often too small to account for the geometry of the solar system. In addition, the high concentration of massive - and therefore very luminous - stars drives additional disc destruction. The combination of both effects,  makes it unlikely that the solar system formed in a starburst cluster.  \\

The central results are:

\begin{itemize}
\item The solar system most likely formed in a leaky cluster environment.
\item As a consequence, the solar birth cluster dispersed  to a large degree over timescales of $\sim$ 20 Myrs, where the cluster radius increased approximately like $r \propto t_c^{0.7}$ and the density in the cluster diminished as $ \rho_{l} [\Msun pc^{-3}] = 150 t_c^{-2.6 \pm 0.2}$ leaving behind a cluster that only has at most 10-20\% of its initial mass.
\item Due to the rapid decrease in density in the solar birth cluster, encounters played a role in shaping the solar system only very early on in the cluster development.  At 1.8 Myr the probability for an encounter at an distance of 100 to 1000AU is $\approx$ 30\%. The encounter most likely took place when the cluster was less than 5Myr old,  because by then the likelihood for such an encounter dropped well below 10\%.. If the development to less concentrates profiles happens on timescales $<$10 Myr, than we would overestimate the encounter frequency in the central parts at the latter stages ($>$3Myr) in our simulations. So our results at $>$3Myr can be regarded as upper limits.
This means that cluster older than 5 Myr would be even less likely to experience a solar system destroying encounter.

\item Hyperbolic encounters with low mass stars and parabolic encounters with high-mass stars
are the most common events for solar-type stars in solar system birth clusters. The former mostly happen in the very early stages ($<$2Myr), the latter preferrentially in the timespan of 2-4Myr.
\end{itemize}

Perturbation of the already formed planetary system \citep{malmberg:07,malmberg:11,lestrade:11,parker:11} by encounters did not take place, because having formed in a leaky cluster the encounter probability for the solar system rapidly dropped as the cluster density decreased precipitately with cluster age.  Even if the solar system was part of the remnant cluster after gas expulsion the stellar density would be $<$ 1 \Msun pc$^{-3}$, far to low to make a close encounter likely.

For the same reason the presence of so many massive stars (10-100) in the solar birth cluster did not lead to its destruction. Although the most massive stars only need 3-5 Myr from their formation to their explosion, even if these supernovae were closely packed at the start of the cluster development due to mass segregation, the rapid cluster expansion of the cluster can account for the solar system being spared further disruption by supernovae at later times.

The here investigated encounters within the distance range of 100 to 1000 AU lead not only to a cut-off in the disc but as well to a steepening of the mass distribution inside the remaining disc. Steinhausen et al (2011) showed that this could result in a $r^{2.2}$-dependence as required for the minimum mass solar nebula (Desch 2007).  

In the still embedded phase the stellar density in leaky clusters were still perhaps somewhat higher (Pfalzner 2011, for an alternative view see Parmentier et al (subm)).  If this were the case, possibly the encounter frequency in the embedded phase could have been higher than in the exposed phase. However, star formation proceeds most likely in an accelerated way with most stars forming just before gas expulsion. In this stage the stellar density is only slightly higher than at the start of the exposed phase, so the sun most likely spent only a short time in this slightly denser environment.

In this study we took the earlier results of the most likely periastron of the solar system-forming encounter having been between 100 and 1000 AU at face value. These former studies often considered only encounters between two solar-type stars and/or parabolic orbits.
However, the results obtained here strongly suggest that these assumptions are not necessarily justified. Future work should consider as well hyperbolic encounters and include the entire mass spectrum of encounter partners. This might be especially important in the evaluation of the likelihood of the solar system having formed in a starburst cluster.

In the current study we included the cluster density development in a simplified way, by looking
at simulations of clusters of different densities in virial equilibrium. Obviously leaky clusters are
at that stage in their development not in virial equilibrium, most prominently illustrated by their mass loss. So in future studies it will be essential to include  the gas expulsion phase to treat 
 the cluster expansion in an appropriate way.   

We have recently modelled the cluster expansion of starburst and leaky clusters after gas expulsion. Preliminary results seem to show that the observational data are surprisingly well reproduced by the simulations (Pfalzner \& Kacmarek, in preparation). However, the fact that bound and unbound stars – with the latter having a preferrential direction – occupying the same space might influence the encounter statistics. This needs to be investigated in future work.
In addition, we find indications that the leaky cluster sequence might only be representative for the development of the most massive clusters in the Galaxy ($\geq$ 10$^4$ \Msun). Lower mass clusters, although starting with approximatedly the same cluster size,
develop along a slightly different path, quickly decreasing so much in density that they fall below the detection limit.  If these results are confirmed, the cluster mass range 2 - 10 $\times$ 10$^3$\Msun\ would require to be investigated in the light of these new results again.
However, given the fast decrease in density in these lower mass clusters the time span of a high encounter probability is extremely short - probably in the range of 1Myr or even less.

\acknowledgements
We would like to thank the referee for the very constructive comments.
This research was supported in part by the National Science Foundation under Grant No. NSF PHY05-51164.

\bibliographystyle{apj}

\end{document}